\pdfoutput=1
\documentclass{JINST}
\pdfoutput=1

\usepackage{graphicx}

\title{Multi-Gigabit Wireless data transfer at 60 GHz}

\author{ H. K. Soltveit$^a$$^,$\thanks{Corresponding author.}  , R. Brenner$^b$, A. Sch\"{o}ning$^a$ and D. Wiedner$^a$\\
\llap{$^a$}Physikalisches Institut Universitaet Heidelberg,\\
  Im Neuenheimer Feld 226, 69120 Heidelberg, Germany\\
\llap{$^b$}Uppsala University, Department of Physics and Astronomy\\
Box 516, SE-75120 Uppsala, Sweden\\

E-mail: \email{soltveit@physi.uni-heidelberg.de}}
\abstract{In this paper we describe the status of the first prototype of the 60 GHz wireless Multi-gigabit data transfer
topology currently under development at University of Heidelberg using IBM 130 nm SiGe HBT BiCMOS technology. The 60 GHz band is 
very suitable for high data rate and short distance applications. One application can be a wireless multi Gbps radial data transmission inside the ATLAS silicon strip detector, making a first level track trigger feasible. The wireless transceiver consists of a transmitter and a receiver. The transmitter  includes an On-Off Keying (OOK) modulator, an Local Oscillator (LO),  a Power Amplifier (PA) and a BandPass Filter (BPF). The receiver part is composed of a BandPass- Filter (BPF), a Low Noise Amplifier (LNA), a double balanced down-convert Gilbert mixer, a Local Oscillator (LO), then a BPF to remove the mixer introduced noise, an Intermediate Amplifier (IF), an On-Off Keying demodulator and a limiting amplifier. The first prototype would be able to handle a data-rate of about 3.5 Gbps over a link distance of 1 m. The first simulations of the LNA  show that a Noise Figure (NF) of 5 dB, a power gain of 21 dB  at 60 GHz with a 3 dB bandwidth of more than 20 GHz  with a power consumption 11 mW are achieved. Simulations of the PA show an output referred compression point  P1dB of 19.7 dB at 60 GHz.}


\begin{document}

\section{Introduction}
The data transfer rate from highly granular tracking detectors  in High Energy Physics (HEP) are limited today by the available bandwidth in the readout links what prevents the detectors to be used for fast triggering. The bandwidth needed to readout all 1-2 hit clusters in the upgraded ATLAS silicon micro-strip tacker is between 50-100 Tb/s. Therefore to get the tracker to contribute to the fast trigger decision, the data transfer bandwidth from the tracker has either to be increased for all data to read out in real time or the quantity of the data has to be reduced by filtering the data or a combination of the two \cite{1}. Our proposal is to readout the data radially, which will help the track trigger and also reduce the latency. The need for higher data rates and more bandwidth has resulted in an increased interests in millimeter wave systems. The 60 GHz unlicensed frequency band available since 2001 is of particular interest for indoor point-to-point multi-gigabit data transfer due to its extremely large amount of spectral bandwidth (7-9 GHz). With such a bandwidth available and the optimum choice of modulation scheme, it would be possible to  achieve a data rate in the 10's Gbps range, and it could therefore be a suitable method to solve the data transfer rate problem. The signal transfer is Line-Of-Sight (LOS), due to indoor use and no obstacles between the layers, short and virtually interference free data transfer distance and narrow antenna beam,  result in secure communication, low power and form factor due to the short distance and high carrier frequency, respectively. These requirements relaxe the specifications substantially. Another attractive feature of the 60 GHz band is the high attenuation through silicon  \cite{1} .
This isolation reduces interference to other users, increasing the frequency re-use and improving the system capacity for short range applications. We are currently working with the design of  a wireless transceiver operating in the 60 GHz band that is capable of delivering multi-gigabit per second data rates, and therefore it could be a potential solution to this challenge. The targeted data rate for our first prototype is 3.5 Gbps.
This paper describes the design of the wireless readout chip designed in IBM 130 nm SiGe HBT BiCMOS technology with an ft/fmax = 200/230 GHz.
In section \ref{sec:over}, the proposed transceiver architecture is presented and the different building blocks are described.  
In section~\ref{sec:tech} the choice of technology is explained. In section~\ref{sec:3D}  the 3D potential is explained.
The conclusions are presented in section~\ref{sec:conclu}.

\section{System Architecture overview}
\label{sec:over}
\begin{figure}[hb]
\begin{center}
\hspace*{-0.0cm}\includegraphics[width=0.6\textwidth,angle=-90]{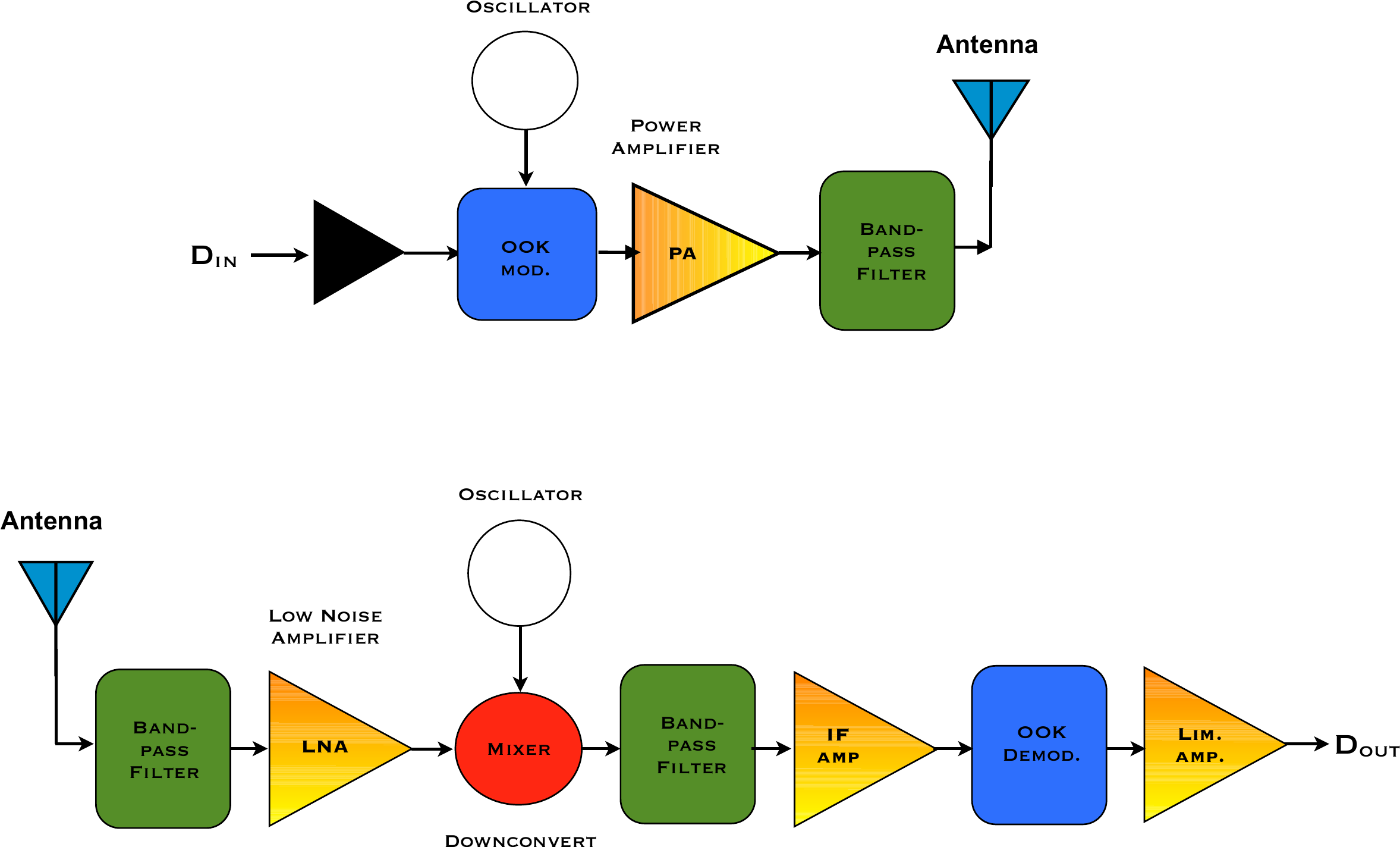}
\caption{Block diagram of the Transceiver. The transmitter is shown at the top and the receiver at the bottom.}
\label{fig:blockpasa}
\end{center}
\end{figure}

The block diagram of the proposed 60 GHz  transceiver chain, is illustrated in Fig.~\ref{fig:blockpasa}. It consist of a transmitter and a receiver part. The transmitter includes a 60 GHz Voltage Controlled Oscillator (VCO), and an ON-OFF Keying (OOK) modulator, and a Power Amplifier (PA). In order to make the data pass through the air, it is modulated onto the carrier signal of 60 GHz.
The PA provides the 60 GHz modulated signal with the required power and amplification.
The PA is followed by a BPF  to suppress the induced broad band noise from the PA itself and carrier feedthrough products from the up-convert process. The filtered signal is then transmitted through the antenna.
The receiver chain is composed of a bandpass-filter, a LNA, a double-balanced mixer, an IF amplifier, an OOK demodulator and a limiting amplifier. The purpose of the receiver is to detect the normally very small MM-wave signal and down-convert it with minimum added noise. The signal from the antenna is first passed through the BPF before it reaches the LNA. The purpose of the BPF is to attenuate out of band interferences. The main function of the LNA is to amplify the weak signal while adding as little noise and distortion as possible. The mixer down-converts the passband signal to a lower frequency, in this case 5 GHz, to make it easier for the following gain, filtering and demodulation stages.
A bandpass-filter then follows in order to reject the superimposed noise, harmonics and mixer inter-modulated products generated in the down conversion process.
The signal is then further amplified with the use of IF amplifier to increase the dynamic range.
OOK is chosen as modulation scheme for the first prototype. It eliminates the digital interface and baseband circuitry.

\subsection{Low Noise Amplifier}
\label{sec:LNA}
 \begin{figure}[ht]
\begin{center}
\hspace*{-1.0cm}\includegraphics[height=0.65\textheight,angle=90]{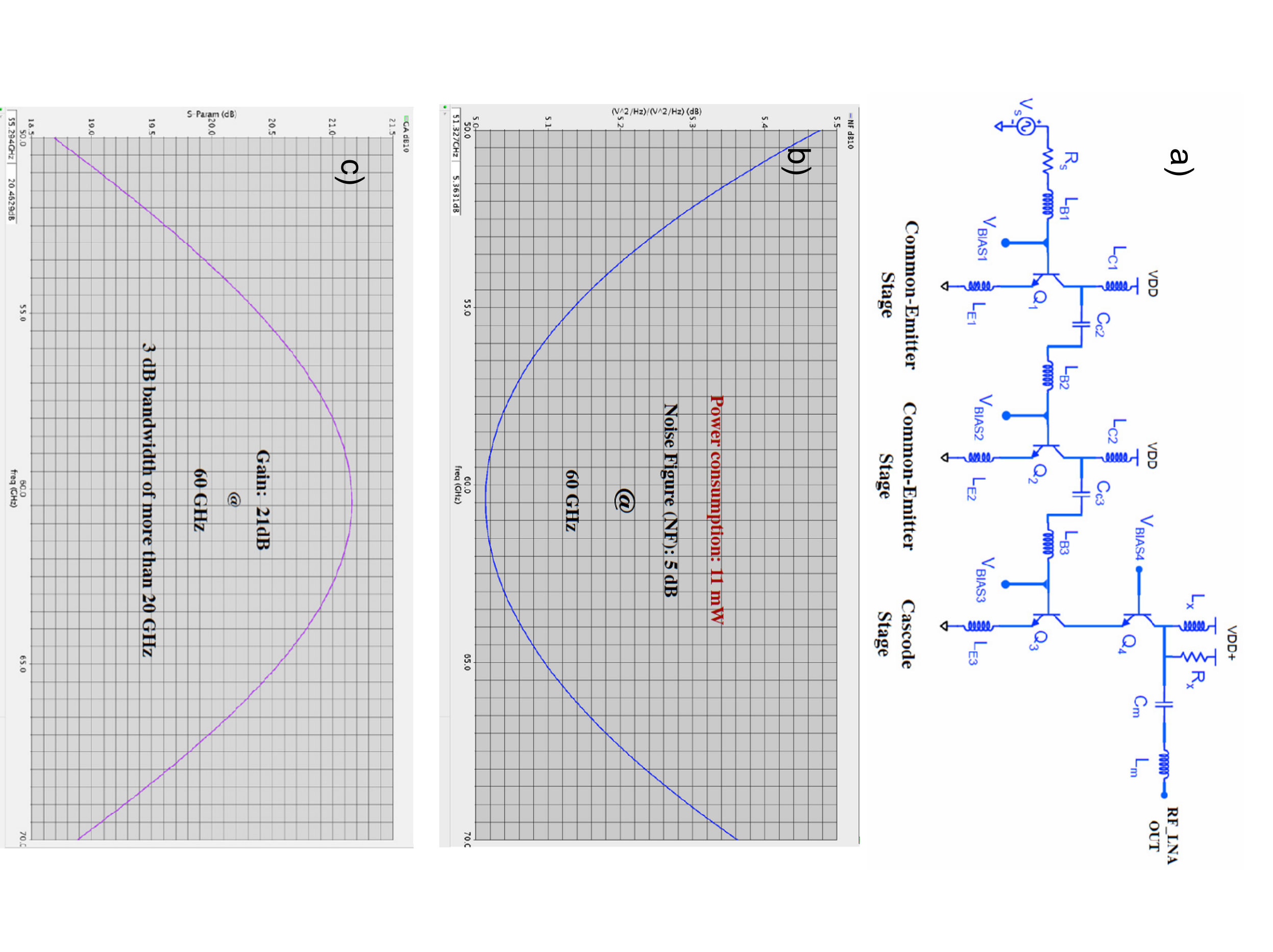}
\caption{a) Block diagram of the three stage LNA. b)  Simulation results an showing NF of about 5 dB at 60 GHz and c) Simulation results showing a power gain of 21 dB and a 3 dB bandwidth of more than 20 GHz with a power consumption about 11 mW.}
\label{fig:Wireless3}
\end{center}
\end{figure}
 The signals power level at the antenna below -56 dBm show the need for low-noise and amplification. The LNA plays a crucial role, as the first active block in the receiver chain following the antenna as its noise figure sets a lower limit on the noise figure of the entire system. So the main goal of the LNA is to amplify the weak signal received on the antenna while adding as little noise as possible. The chosen topology for the LNA is shown in Fig.~\ref{fig:Wireless3}a.  In order to provide  enough gain,  low noise and isolation, the LNA is implemented with a three stage cascade design.  
 \
The inductively generated emitter topology  has been chosen for the two first stages due to its  better gain, noise figure and to improve the linearity.  The two first stages are biased at minimum NF current to minimize the noise, while the last stage provides high gain and isolation.  
 For input matching of the first stage, noise and input matching technique is used simultaneously by tuning the transistor size of Q1, the emitter degeneration inductor LE1, the input capacitor and the inductor.
This gives the best compromise between maximum gain and minimum NF.   
Each stage is biased independently. For simplicity, the bias network is not shown.  
Simulations of the LNA  Fig.~\ref{fig:Wireless3}c give a NF of 5 dB  with a power consumption of 11 mW at 60 GHz. A simulated gain of 21 dB with a bandwidth of 20 GHz is also  achieved as seen in Fig.~\ref{fig:Wireless3}b.

\subsection{Gilbert mixer}
\label{sec:mixer}

\begin{figure}[h]
\begin{center}
\hspace*{-1.0cm}\includegraphics[width=0.49\textwidth,angle=-90]{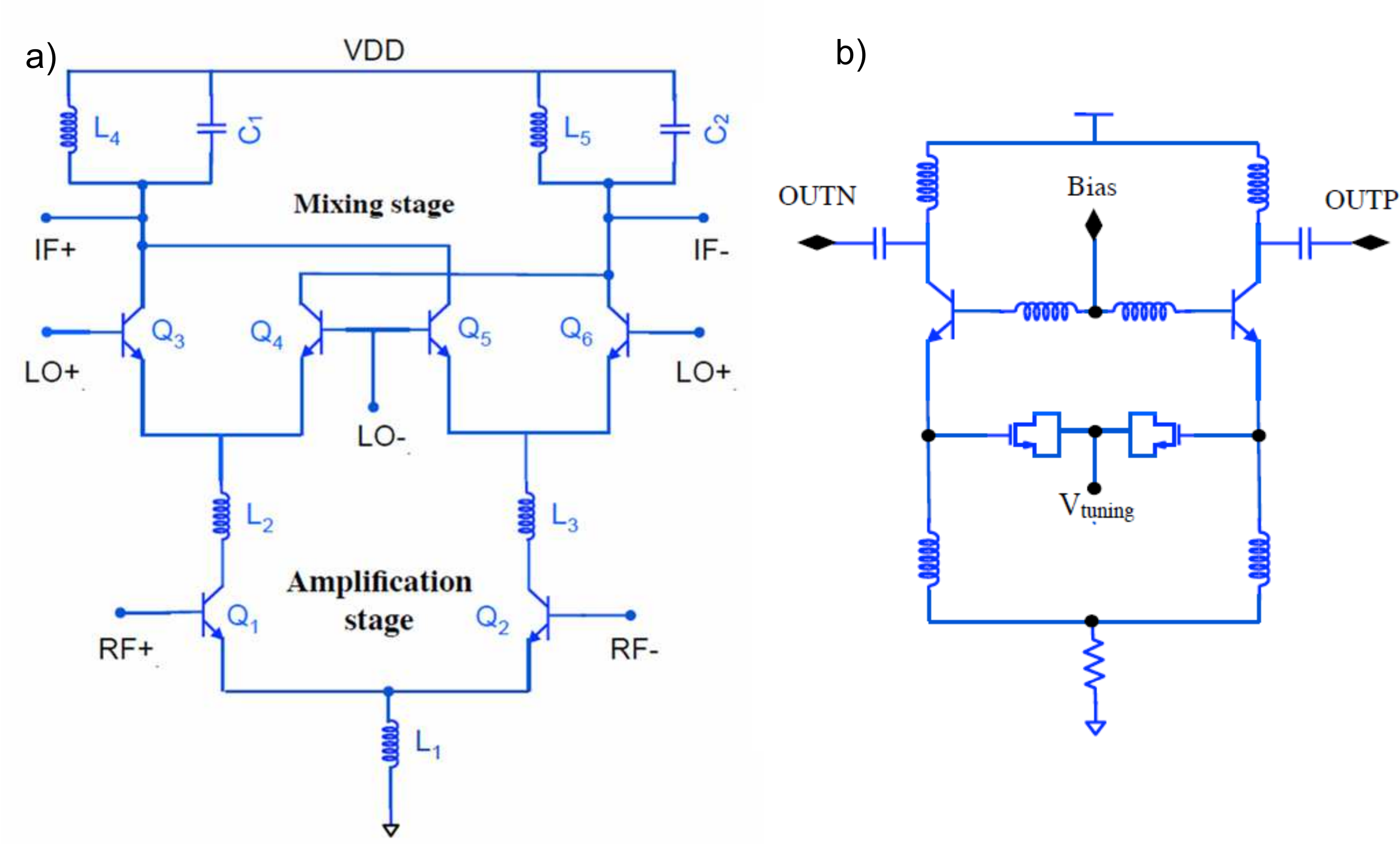}
\caption{A simplified block diagram of the double balanced Gilbert mixer (left) and the Voltage Controlled Oscillator (right).}
\label{fig:Wireless7-1}
\end{center}
\end{figure}

The mixer performance has a great influence on the characteristics of the overall front end. Since it follows the LNA, the issue of linearity becomes significant because it must handle amplified signals. 
While it may seem that the issue of noise is relaxed by the LNA gain, in practice, mixers exhibit a high noise figure that should also be taken into account. The gain of the mixer is also important to compensate for the intermediate frequency filter loss and to reduce the noise contribution from the IF stages. So in a mixer  design, it is always necessary to achieve a trade-off among the three main parameters: gain, noise figure and linearity. A double balanced Gilbert cell is chosen as a mixer Fig.~\ref{fig:Wireless7-1}a. It takes the input and  down-converts it to 5 GHz for easy signal processing in the receiver. Strong LO-IF feedthrough is suppressed by the double balanced mixer. All the even harmonics are cancelled. All the odd harmonics doubled. It provides a reasonable gain. 

\subsection{Voltage controlled oscillator}
\label{sec:VCO}
The voltage controlled oscillator (VCO) is a key component in transceivers. The voltage controlled frequency of operation is achieved by voltage dependant capacitance devices such as varactors. The VCO provides the reference frequency to modulate/demodulate the RF signal. VCO's are particularly sensitive to phase noise, which physically represents the up-conversion of low frequency (1/f) noise to RF frequency. The Colpitts topology in  Fig.~\ref{fig:Wireless7-1}b is chosen for low- phase noise and high frequency behavior. 
For this prototype design, the goal of the VCO is to have a tuning range of 7 GHz, and a phase noise of less than -90 dBc/Hz at 1 MHz.

\subsection{Power amplifier}
\label{sec:PA}
\begin{figure}[ht]
\begin{center}
\vspace*{-0.9cm}
\hspace*{-1.0cm}\includegraphics[height=0.45\textheight,angle=0]{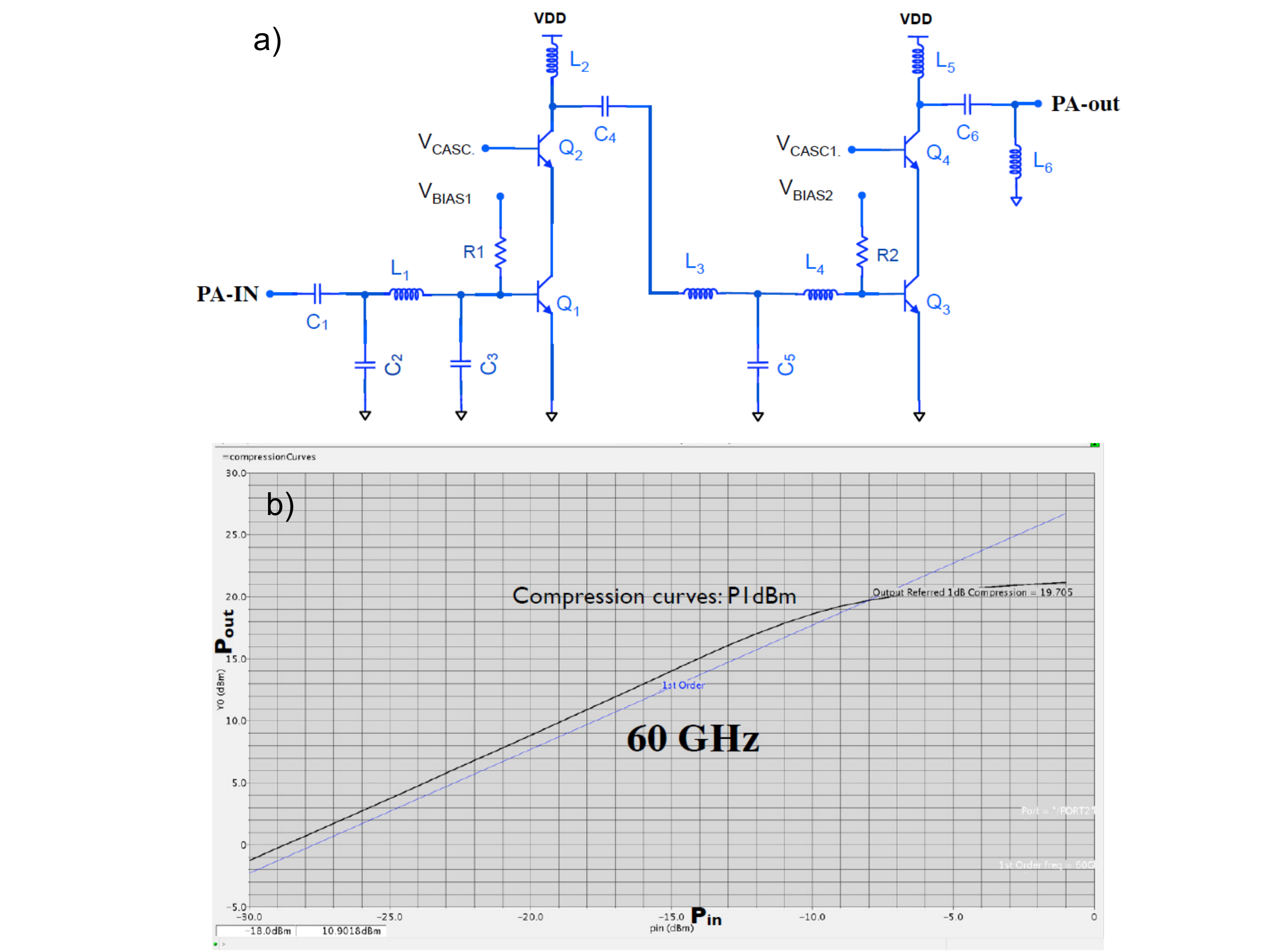}
\caption{A simplified block diagram of the Power Amplifier a) and simulation of its compression point  b). }
\label{fig:PA}
\end{center}
\end{figure}

The power amplifier is the final active block in the transmit path of the transceiver. The purpose is to provide the signal with the required power level to ensure that the transmitted signal reaches the receiver with the required power level.
 In order to provide high gain, that improves the Power Added Efficiency (PAE) and isolation, a two stage cascode topology was chosen as shown in Fig.~\ref{fig:PA}a.  The PA is biased in class AB mode to maintain high efficiency.
The low external base resistance minimizes impact ionization and thus maximizes the breakdown voltage of the output device.
Preliminary simulations of the PA gives a compression point of about 20 dBm for a input of about 8dBm. The input power return loss is about -65 dB.
Since the need for extreme transmission power is not required in our specific design a new PA with less transmitting power (0 dBm) is under development. This will reduce the power consumption of the whole system drastically, since the PA is the main contributor.

\subsection{On-Off Keying}
\label{sec:OOK}
On-off shift keying (OOK) modulation/demodulation scheme will be used for the first prototype \cite{2}. Because of the relative low complexity of implementation and very high data rates offered. The OOK represents the digital data as the presence  as "one" or absence "zero" of a carrier wave. OOK modulation has the advantage of allowing the transmitter to idle during the transmission of a "zero", and therefore saving power. Other advantages are the need of non-linear amplifiers, immunity to VCO phase noise and no need of a DSP circuitry.
OOK supports data rates up to half the available bandwidth \cite{2}, but with an available bandwidth of 7 GHz, a data rate of 3.5 Gbps is possible.

\section{Link Budget and channel characterization}
\label{sec:link}
A link budget evaluation is mandatory to get insight into the specifications for  some of the most important blocks. To decide the appropriate system design at 60 GHz a characterization of the channel is done. Important specifications are range limitations, material propagation,  available and permitted output power, available bandwidth,  receiver sensitivity, multi-path propagation and environmental conditions.
A loss from the feed-line of 4 dB, the passive filters of 6 dB, and a path loss of 60 GHz signal is 68 dB for a distance of 1 meter. 
The required SNR is 17 dB for a BER of 10$^{-12}$. A total antenna gain of 16 dBi, fade-margin of 20 dB and a bandwidth of 5 GHz is here anticipated.
For a maximum LOS distance of about 1m a, transmit power of about 20 dBm is required to fulfill these requirements.
 
\section{Choice of Technology}
\label{sec:tech}
The technology chosen must be able to fulfill requirements such as noise and linearity and at the same time have a high production yield at a reasonable cost. Considering availability, cost and the low integration level of these technologies, only the two mainstream technologies CMOS and SiGe HBT BiCMOS come into question. 
Compared to CMOS,  SiGe HBT transistors have less process variability and/or better yield since its performance depends on the vertical diffusion, that is less variable than the CMOS lateral gate length, 
higher carrier mobility constants and higher breakdown voltages. 
The modeling for SiGe is easier for high frequency, which turns into higher probability for first time pass. 1/f noise is much  lower in SiGe HBT's.
Bipolar transistors have higher gm/I, which means more gain for the same bias current,  that can be traded off to improve the 
NF.  The SiGe BiCMOS technology combines both high speed HBTs with relatively high breakdown voltage and standard CMOS transistors allowing a very high integration level. The electrical performance has been tested after irradiation and it was concluded that the technology is suitable for  condition of the Super-LHC upgrade  \cite{3}. Based on these reasons, it was decided to design the 60 GHz transceiver chain in the 130 nm SiGe HBT BiCMOS technology.

\section{3D development}
\label{sec:3D}
One of the biggest challenges when designing transceivers is the non optimal implementation of antenna and passives in silicon due to the high loss through the silicon substrate. A natural step could be to go to 3D.
The key benefit of 3D heterogenous designs with Through Silicon Vias (TSV) is its ability to combine optimum, mature and incompatible process technologies most suitable for the specific function, e.q., logic, memory, analog and RF circuits in specialized processes.
This keeps development cost low, minimizes risk and time.
TSV offer significant power savings and much better performance gains than any other connecting stacked die. Also, they do not require high drive I/Os, saving not only power, but also die area and silicon cost. TSVs enable die stacking on the wafer level and produces hundreds of thousands of die stacks simultaneously. As manufacturing techniques mature, TSVs promise to add more cost savings to the compelling technical benefits  \cite{4}.

\section{Conclusions and further work}
\label{sec:conclu}
A transceiver using the 60 GHz band as a solution to the bandwidth limitation for a first level track trigger and to meet the requirement set by the high data rate in HEP has been proposed. 
Individual blocks under development designed in 130 nm SiGe HBT BiCMOS  have been described. First simulations show that reasonable noise and power levels at 60 GHz is achieved.  The  choice of technology and the benefit of going 3D has been briefly discussed. The first prototype of the transceiver is forseen to be submitted mid 2013.

\end{document}